\documentclass{article}

  \usepackage[final]{neurips_2020}

\usepackage[utf8]{inputenc} 
\usepackage[T1]{fontenc}  
\usepackage{hyperref}    
\usepackage{url}      
\usepackage{booktabs,graphicx,amsmath,float}    
\usepackage{amsfonts}    
\usepackage{nicefrac}    
\usepackage{microtype}   
\usepackage{epstopdf}
\setcitestyle{numbers,square}
\title{Learn to Sing by Listening: Building Controllable Virtual Singer by Unsupervised Learning from Voice Recordings}

\author{%
 Wei Xue$^1$, Yiwen Wang$^2$, Qifeng Liu$^{3,1}$, Yike Guo$^1$\\
 $^1$ Hong Kong University of Science and Technology, $^2$ Hong Kong Chu Hai College, \\
 $^3$ Hong Kong Institute of Science \& Innovation, Chinese Academy of Sciences
}

\begin{document}

\maketitle

\begin{abstract}
The virtual world is being established in which digital humans are created indistinguishable from real humans. Producing their audio-related capabilities is crucial since voice conveys extensive personal characteristics. We aim to create a controllable audio-form virtual singer; however, supervised modeling and controlling all different factors of the singing voice, such as timbre, tempo, pitch, and lyrics, is extremely difficult since accurately labeling all such information needs enormous labor work. In this paper, we propose a framework that could digitize a person's voice by simply ``listening'' to the clean voice recordings of any content in a fully unsupervised manner and predict singing voices even only using speaking recordings. A variational auto-encoder (VAE) based framework is developed, which leverages a set of pre-trained models to encode the audio as various hidden embeddings representing different factors of the singing voice, and further decodes the embeddings into raw audio. By manipulating the hidden embeddings for different factors, the resulting singing voices can be controlled, and new virtual singers can also be further generated by interpolating between timbres. Evaluations of different types of experiments demonstrate the proposed method's effectiveness. The proposed method is the critical technique for producing the AI choir, which empowered the human-AI symbiotic orchestra in Hong Kong in July 2022.
\end{abstract}

\section{Introduction}
We are entering a new era in which the boundary between real and virtual worlds is increasingly blurred, eliminating the geographical barriers between people and the gaps between humans and AI. This further facilitates co-inspiring and co-creation between humans and AI to push the boundaries of science and art. Digital humans can be created indistinguishable from real humans; generating natural and personalized voices is essential since the voice conveys not only the content information for communication but also personalized information such as timbre, accent, and cadence. In this paper, we aim to produce audio-form digital humans capable of singing, i.e., virtual singers, with wide-ranging applications in entertainment, virtual assistants, cultural preservation, and digital immortality.

Creating natural voices from the machine is conventionally tackled by the problem of text-to-speech (TTS), which syntheses speaking speech waveforms according to the text specifying the content. Early TTS approaches seek to estimate the over-simplified linear filters modeling the physical structure of vocal organs \cite{Thyssen1994,Markel1976,Rabiner2010}, and current mainstream approaches train the deep neural networks (DNNs) in a supervised manner to model the speech signals for different contents and dynamics while preserving the timbre. Typically, these approaches first utilize an acoustic modeling network to transform the text into the time-frequency mel-spectrogram \cite{Wang2017,Shen2018,Yu2020,Vasquez2019,Ren2019,Ren2021}, and then adopt a vocoder \cite{Oord2016,Oord2018,Mehri2017,Valin2019} to transform the mel-spectrogram into the time-domain waveforms. Widely used acoustic modeling networks include Tacotron \cite{Wang2017}, DurIAN \cite{Yu2020}, and FastSpeech \cite{Ren2019,Ren2021}, and vocoders include WaveNet \cite{Oord2016}, HiFi-GAN \cite{Kong2020} and MelGAN \cite{Kumar2019}. To achieve singing voice synthesis (SVS), the lyrics and melody information are jointly used as the input of the acoustic modeling network \cite{Lu2020,Song2022,Liu2022,Huang2022}, and vocoders similar to the TTS are adopted to produce the audio signals finally.

A major problem for supervised SVS is that a large annotated dataset is required, which includes the audio and the corresponding scripts indicating the content and melody. Although there are some public SVS datasets such as OpenCpop \cite{Wang2022} and VocalSet \cite{Wilkins2018}, building such datasets needs extensive labor work for recording and annotating. This hinders flexibly building the voice models for an arbitrary person and also prevents simulating the singing skills of highly professional singers because of the cost of inviting singers for recording. To eliminate the reliance on annotation, singing voice conversion (SVC) can be used to essentially perform style transfer on audios, where the timbre is defined as the ``style''. Many methods \cite{Chou2018,Chou2019,Wang2018a,Lu2019,Wang2021} are proposed to disentangle the content and timbre information with unsupervised learning and then replace the timbre with the target speaker. However, as only the timbre is replaced with the target singer, it is difficult to model the unique singing skills of the singer, and precisely controlling other diversified singing characteristics is also not straightforward.

In this paper, we propose a new framework to digitize the voice in a fully unsupervised manner. The proposed method could simply rely on the audio recordings of any content and language to build a flexible voice model and make it possible to control the detailed characteristics such as singing pitch, melody, and lyrics. In this way, the proposed method can even generate singing from datasets of speaking voices.

A variational auto-encoder (VAE) based framework is developed, which leverages a set of pre-trained models to encode the audio as various hidden embeddings representing different factors of the singing voice, and further decodes the embeddings into raw audio. By manipulating the hidden embeddings for different factors, the resulting singing voices can be controlled, and new virtual singers can also be further generated by interpolating between timbres. Furthermore, by training on large-scale data, the model can also learn to model other unique skills, including the accent and emotional expression of the singer. We conduct experiments on different datasets for various tasks, and the results demonstrate the effectiveness of the proposed method. The proposed method is also the key technique for producing the AI choir, which empowered the human-AI symbiotic orchestra in Hong Kong in July 2022 \cite{Guo2022}.

The rest of this paper is organized as follows. In Section 2, we review the related works. The proposed framework for unsupervised voice modeling will be introduced in Section 3, including the encoder and decoder models, as well as the end-to-end adaptation. How to manipulate the representations to achieve controllable singing will also be described. In Section 4, we explain how to generate the AI choir based on the proposed framework. Experimental results will be shown and discussed in Section 5.

\section{Related Work}
\subsection{Speech Synthesis}
A conventional way to digitize the human voice is speech synthesis, or TTS, which generates natural speech according to the text inputs. The state-of-the-art TTS systems are based on neural networks to model the complicated dynamics of natural speech, and datasets with pairwise audios and texts are used to train the models in the supervised learning scheme. The pipeline of TTS generally consists of an acoustic model which converts the textual information into the audio mel-spectrogram and a vocoder that further generates the audible waveform. The acoustic model mainly focuses on learning the low-level speech representations from text, and the vocoder aims at generating ultra-long signals (e.g., 24,000 samples per second) with high fidelity \cite{Oord2018}.

Early DNN-based approaches for acoustic modeling generally use RNN to model the temporal dependencies of speech, such as Tacotron \cite{Wang2017} and DurIAN \cite{Yu2020}, and produce the mel-spectrogram in an autoregressive way. Transformer-based methods, including \cite{Li2019,Chen2020,Ren2019,Ren2021}, are then proposed, significantly speeding up the generation process by parallel computation in the attention. Variational autoencoder (VAE) is also used for acoustic modeling, typical methods include GMVAE-Tacotron \cite{Hsu2019} and VAE-TTS \cite{Zhang2019}. Recently, diffusion-based acoustic models, such as \cite{Jeong2021,Popov2021}, have been further developed to improve the quality of synthesized speeches. When using the acoustic model for SVS  \cite{Lu2020,Song2022,Liu2022,Huang2022}, the main modification is that the pitch and duration of the lyrics are explicitly given rather than being estimated by separate modules in the speaking speech synthesis.

The vocoder is typically regarded as a sequence-to-sequence modeling problem. An important work is WaveNet \cite{Oord2016}, which uses the autoregressive (AR) convolutional neural network for sequence modeling. Parallel WaveNet \cite{Oord2018} further uses the inverse autoregressive flow (IAF) to distill the knowledge from a pre-trained WaveNet teacher, substantially improving the inference speed. GAN-based methods are also proposed to improve speech quality, including MelGAN \cite{Kumar2019}, GAN-TTS \cite{Binkowski2020}, and HiFi-GAN \cite{Kong2020}. Diffusion models further improve speech quality, and typical works include SpecGrad \cite{Koizumi2022} and DiffWave \cite{Kong2021}. Differential signal processing is also used to design the vocoders, for instance, Source-Filter HiFi-GAN \cite{Yoneyama2022} and SawSing \cite{Wu2022}.

Separately optimizing the acoustic modeling and vocoder helps to improve the stability of model training, and the universal vocoders can also be trained to synthesize voices of different timbres. Nevertheless, many end-to-end methods have been developed to produce the audio waveforms directly from the textual inputs. Famous works include Char2Wav \cite{Sotelo2017}, ClariNet \cite{Ping2019}, and FastSpeech 2s \cite{Ren2021,Lim2022}, which generally combine acoustic modeling and vocoder into a large encoder-decoding framework.

\subsection{Voice Conversion}

The significant reliance on the large and carefully annotated dataset is a primary obstacle to exploiting speech synthesis to digitize the human voice. The dataset includes textual information about the content of speaking, and for SVS, the duration, and pitch of each word in the lyrics should be additionally precisely given. Several hours of training data are needed to produce high-quality audio. Such a requirement significantly increases the difficulty of training a voice model for the ordinary person since specialized voice recording and annotation are costly. For SVS, the singing skills of the produced voice models are also limited since inviting many pop stars for recording is nearly impossible.

By applying style transfer, which is extensively studied in computer vision \cite{Gatys2015,Gatys2016,Li2017,Zhu2017,Choi2018}, to the audios, the problems of voice conversion (VC) and SVC are formulated, which treats the timbre as the style of audio and the rest information as the content. In this way, inspired by methods such as CycleGAN \cite{Zhu2017} and StarGAN \cite{Choi2018} for images, unsupervised learning can disentangle the hidden audio representations on nonparallel data from different people. Further, VC can be achieved by changing the timbre embedding, which yields CycleGAN-VC \cite{Kaneko2018}, StarGAN-VC \cite{Kameoka2018}, and StarGANv2-VC \cite{Li2021}. The main focus has been disentangling linguistic features from the speaker representation, such as the VAE-based methods \cite{Lu2022,Williams2021}, phonetic posteriorgram (PPG)-based methods \cite{Li2021a,Sun2016}, as well as the vector quantification (VQ) based methods \cite{Tang2022,Wang2021,Wu2020}. Although these methods can change the timbre of the generated audio to the target, they cannot fully express other unique characteristics of the human voice, such as accent, emotion, and rhythm. On the other hand, although many unsupervised methods have been developed to disentangle the audio representations, explicitly disentangling different characteristics remains challenging.

\section{Proposed Framework for Voice Digitization}
\label{proposed}

\begin{figure}
 \centering
 \includegraphics[width=13cm]{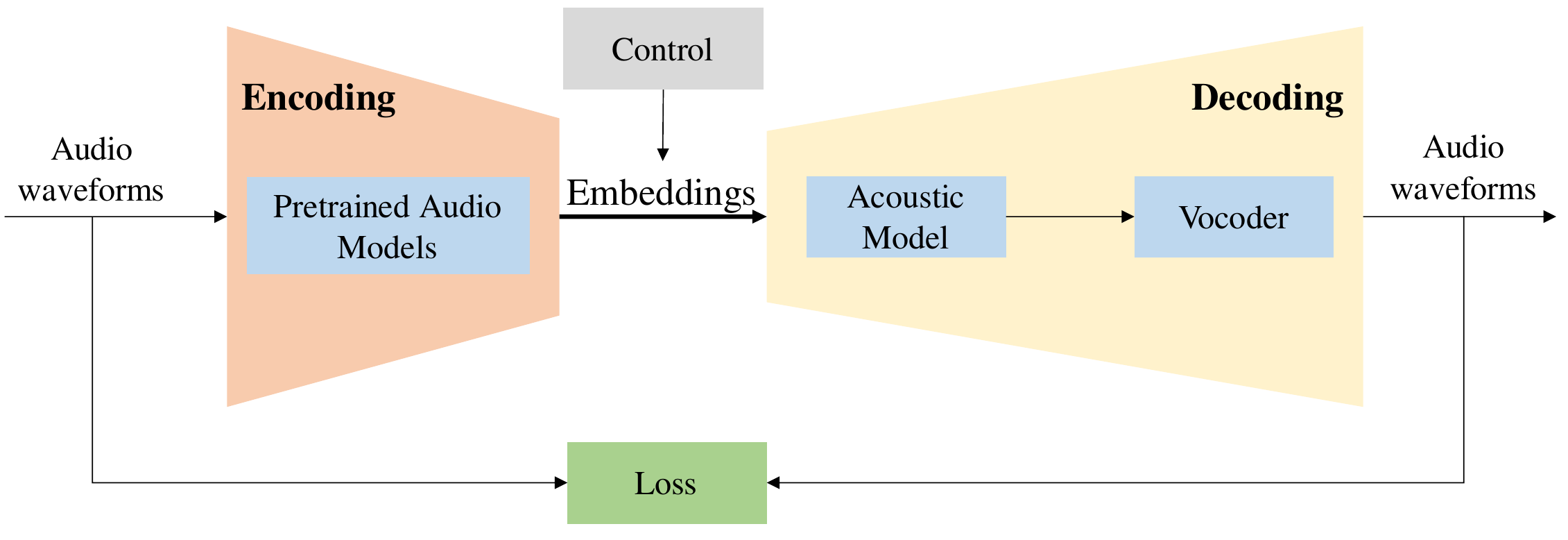}
 \caption{Diagram of the overall framework.}
 \label{figvae}
\end{figure}

In this paper, we propose a new method that can digitize anyone's voice based on their recordings without annotation. The recordings can be of arbitrary content and language, and the proposed method can predict singing voices only from speaking recordings. As shown in Fig.~\ref{figvae}, an unsupervised VAE-based framework is developed. Similar to humans using existing skills to learn new tasks, we largely leverage the skills of a set of pre-trained audio models to decouple different characteristics of sounds. Similar to the pipeline of TTS, the proposed framework consists of an acoustic model which maps different controlling factors to the mel-spectrogram and a vocoder that converts the mel-spectrogram to audio waveforms. After training the acoustic model and vocoder separately, the end-to-end training is further adopted to finetune the two networks. Details of the proposed framework will be described below.

\subsection{Pre-training Audio Models}

To identify detailed audio characteristics, we aim to figure out a) what is the audio content, b) who is speaking or singing, and c) what is the melody of the speaking or singing. We note that training data for the target speaker containing pairs of audio and textual information is not available. Therefore, instead of performing end-to-end unsupervised learning to disentangle audio representations, we rely on models trained on other annotated datasets to identify distinctive characteristics. In this way, the reliance on an annotated target speaker dataset is eliminated, while discriminative embeddings can be generated.

\begin{figure}
 \centering
 \includegraphics[width=8cm]{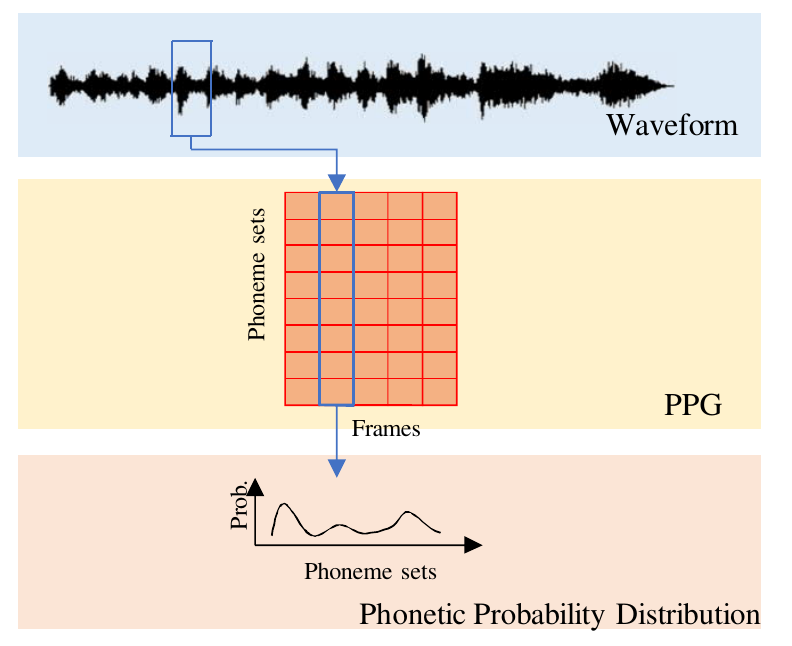}
 \caption{Illustration of PPG. Each column of the two-dimensional PPG represents the probability distribution over different phonemes.}
 \label{fig1:ppg}
\end{figure}

The automatic speech recognition (ASR) model trained on the large-scale ASR dataset can identify the content information. The Wenetspeech dataset \cite{Zhang2022a}, consisting of over $10,000$ hours of accurately labeled Mandarin speeches, is used for training, and the pre-trained model \cite{Wenet2022} based on the U2++ network \cite{Zhang2022b} is utilized to generate the audio content embedding, which is the output of the Conformer \cite{Gulati2020} encoding block. As shown in Fig.~\ref{fig1:ppg}, the audio content embedding is essentially the two-dimensional PPG, and each column indicates the probability distribution of the current frame on different phones. When feeding the unannotated audio from the target speaker to the pre-trained ASR model, its content information can be represented by the PPG. The frame size for the ASR processing is $25$~ms, and the hop size is $10$~ms. Assuming $T$ frames are contained in the utterance, to accelerate the decoding, $T/3$ PPG frames are obtained with the subsampling factor as $3$ in the ASR decoding. The resulting PPG is then up-sampled by $3$ times to match the original number of frames.

Similarly, the speaker identity is obtained using the VoxCeleb2 \cite{Chung2018} dataset for speaker identification. An ECAPA-TDNN model \cite{Desplanques2020} pre-trained using the SpeechBrain \cite{Ravanelli2021} is adopted to generate the speaker embeddings, which are the outputs of the model. For one utterance, only one speaker embedding is obtained, and the embeddings of all utterances for the same speaker are averaged to finally represent the timbre. The resulting embedding is further replicated over $T$ times to represent the identity of each frame.

We finally identify the melody of the speech and singing voice by estimating the pitch contours in the voiced audio. The pre-trained CREPE model \cite{Kim2018} is exploited, and the resulting contour for the $T$-frame utterance is a $T\times1$ signal, indicating the pitch of each frame.

\begin{figure}
 \centering
 \includegraphics[width=11cm]{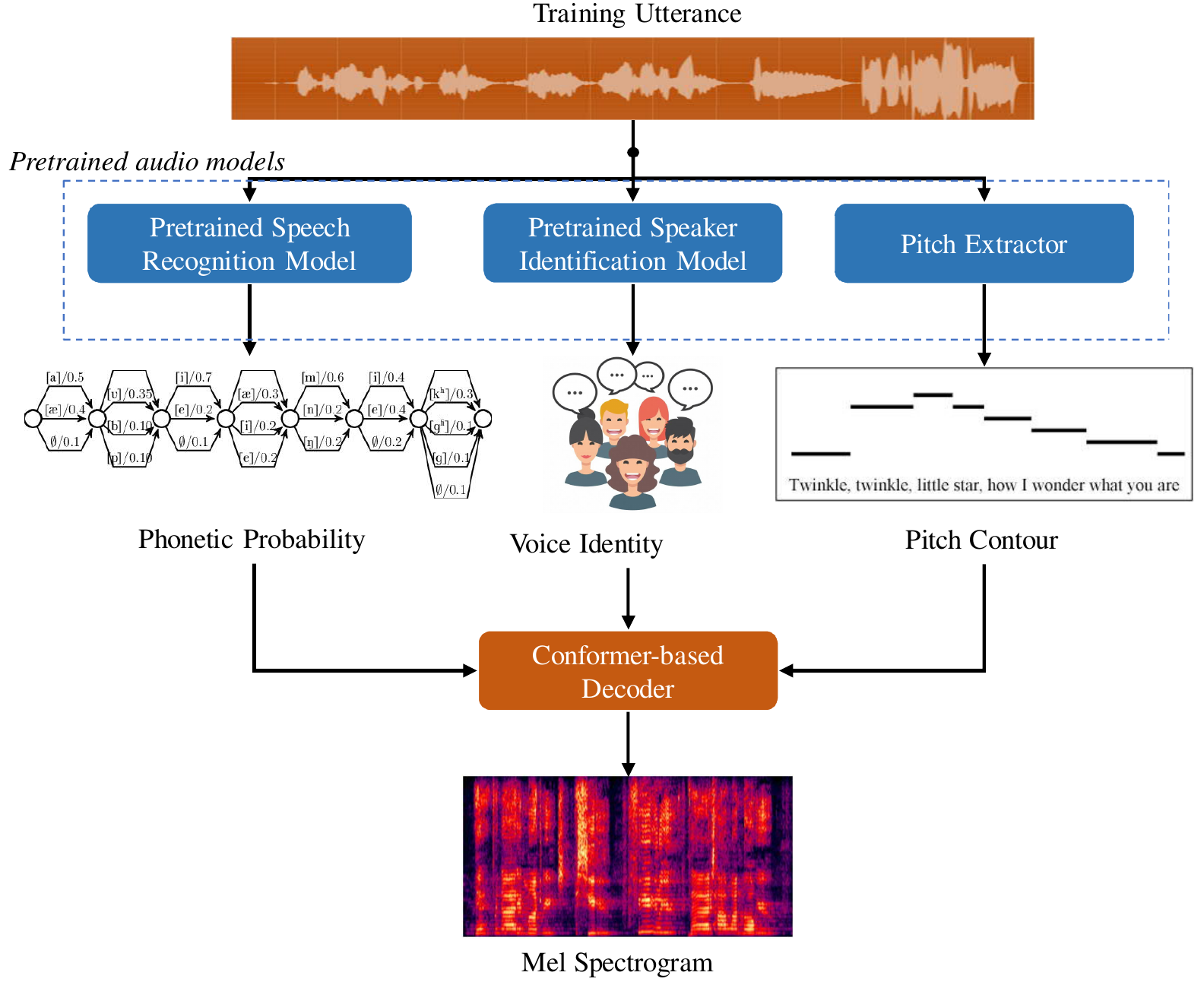}
 \caption{Diagram of the acoustic model, which converts embeddings extracted from pre-trained audio models to the mel-spectrogram of the input signal.}
 \label{fig2:acoustic_model}
\end{figure}

\subsection{Acoustic Model}

With pre-trained audio models, we can convert the audio signal without any annotation into a set of embeddings representing the different characteristics. An acoustic model is further trained to convert the obtained embeddings into the mel-spectrogram, which will be used to synthesize audible waveforms later. Since the mel-spectrogram has much richer information than the embeddings from pre-trained models, the acoustic model learns to represent the uniqueness of the speaker, thereby digitizing the person's voice. In addition, the embeddings also provide the interface to control the resulting mel-spectrogram.

The diagram of the proposed acoustic model is shown in Fig.~\ref{fig2:acoustic_model}. Given the extracted embeddings representing the content, identity, and melody information, a Conformer-based decoding network is designed to reconstruct the mel-spectrogram of the input utterance. The mel-spectrogram can be directly computed by applying the Mel filterbanks to the input signal in the short-time Fourier Transform (STFT) domain. Therefore, the acoustic model can be seen as an auto-encoder, and the training can be conducted fully unsupervised.

\begin{figure}
 \centering
 \includegraphics[width=13cm]{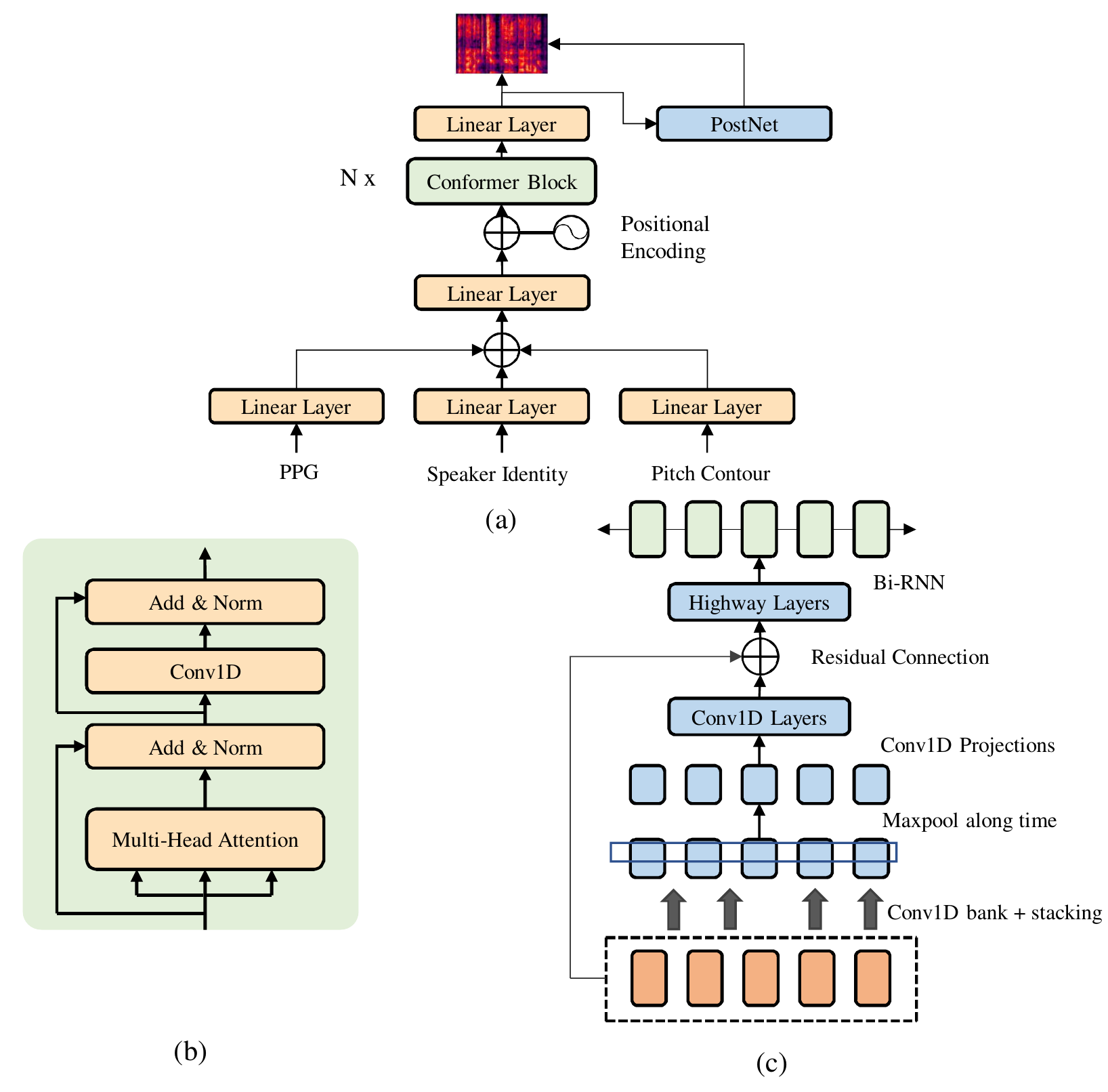}
 \caption{Details of the Conformer-based decoder in the acoustic model. (a) Overall structure; (b) Conformer Block; (c) PostNet.}
 \label{fig3:conformer}
\end{figure}

The details of the Conformer-based decoder are illustrated in Fig.~\ref{fig3:conformer}. In Fig.~\ref{fig3:conformer} (a), three linear layers first process different audio embeddings separately to ensure all embeddings have the same dimension. Then, the projected embeddings are summed up and further transformed by a linear layer before being fed into the Conformer block. The acoustic model finally predicts the mel-spectrogram of the utterance producing the embeddings, and similar to \cite{Ren2019}, a PostNet is used as an auxiliary network to improve the prediction performance.

The Conformer block is widely used for ASR and TTS \cite{Gulati2020,Ren2019,Chang2023}, which basically integrates a convolution block into the Transformer, and its structure is shown in Fig.~\ref{fig3:conformer} (b). The structure of the PostNet is shown in Fig.~\ref{fig3:conformer} (c), which adopts the convolutional RNN (CRNN) with skip connections to model the temporal and structural information of the mel-spectrogram.

During training, only the Conformer-based decoder is optimized using the combination of $L_1$ and $L_2$ norms:
\begin{align}
&L(\mathbf{S}_{\textrm{linear}},\mathbf{S}_{\textrm{PostNet}},\mathbf{S}_{\textrm{gt}}) \nonumber\\
&= ||\mathbf{S}_{\textrm{linear}}-\mathbf{S}_{\textrm{gt}}||_1+||\mathbf{S}_{\textrm{linear}}-\mathbf{S}_{\textrm{gt}}||_2+||\mathbf{S}_{\textrm{PostNet}}-\mathbf{S}_{\textrm{gt}}||_1+||\mathbf{S}_{\textrm{PostNet}}-\mathbf{S}_{\textrm{gt}}||_2,
\end{align}
 where $\mathbf{S}_{\textrm{linear}}$,$\mathbf{S}_{\textrm{PostNet}}$ are the mel-spectrograms output by the last linear layer and PostNet, respectively, and $\mathbf{S}_{\textrm{gt}}$ is the ground-truth mel-spectrogram.

\subsection{Vocoder}
\begin{figure}
 \centering
 \includegraphics[width=13cm]{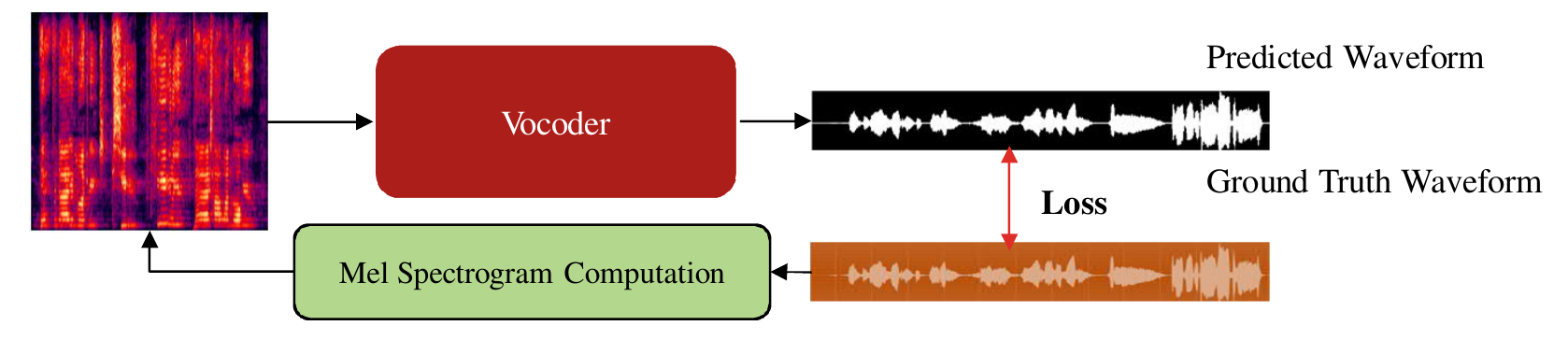}
 \caption{Illustration of the vocoder training.}
 \label{fig4:vocoder}
\end{figure}
The mel-spectrogram is further converted into the audio waveform through a vocoder. Here, the widely used  HiFi-GAN \cite{Kong2020} vocoder is adopted, which uses multi-scale and multi-period discriminators to ensure the fidelity of the produced waveforms. As illustrated in Fig.~\ref{fig4:vocoder}, for each utterance, the ground-truth mel-spectrogram is first computed and is then fed into the vocoder to produce the original signal.

\subsection{End-to-end Training}
We note that in practice, the mel-spectrogram predicted by the acoustic model rather than the ground-truth one will be used to generate the audio signal. However, there will always be an error between the predicted mel-spectrogram and the ground truth, which will finally affect the quality of the produced audio. In Fig.~\ref{figvae}, the whole framework aims to perform end-to-end VAE over the audio waveforms of the target person. Therefore, end-to-end training over the acoustic model and vocoder will be performed to finetune both models further.

In the end-to-end training, the acoustic model and vocoder are combined to transform the embeddings extracted from the pre-trained audio models directly to the raw waveforms, and they are alternatively optimized by freezing the other model when updating the weights. The loss function of the HiFi-GAN is used to optimize the end-to-end framework.

\subsection{Controllable Audio Generation}

Given the acoustic model and vocoder, as shown in Fig.~\ref{figvae}, controllable audio signals can be generated by manipulating the audio embeddings fed into the acoustic model, and we can produce singing voices even from the speaking recordings.

The PPG representing the audio contents can be obtained by applying the ASR model on existing speech recordings with target contents, e.g., a real or synthesized reading speech or the vocal track of a song. Forced alignment (FA) \cite{Zhang2022a} can be conducted to determine the interval of each phoneme, and the PPG corresponding to each phoneme can be re-sampled to adjust to the target duration.

The pitch contour can also be extracted from existing audio which can be either speaking or singing and can be further edited in an explainable way. If the training data has large variations on the pitch, even when only the speaking data is contained, the model can learn how to generate the sounds for the target pitches in a singing melody, such that singing voices can be produced.

We can produce either a personalized model for each person, corresponding to the audio-form digital twin, or a general model that can change the timbres to create non-existing humans. A pair of specialized acoustic models and vocoder can be trained to fully model the uniqueness of an individual's voice, such as the timbre, accent, and subtle rhythm control of speaking and singing. In this case, the speaker identity embedding in can be fixed to a zero-valued vector or be removed in the framework shown in Fig.~\ref{fig2:acoustic_model}. To produce variable timbres, audio data from different people can be jointly used to train the acoustic model, in which case the speaker identity embedding can be manipulated to control the timbre. A universal vocoder can also be trained to convert any mel-spectrogram to an audible signal.

\section{Case Study: AI Choir Generation}

This section explains how to generate an AI choir composed of hundreds of virtual singers based on the proposed method, which empowered the human-AI symbiotic orchestra in Hong Kong in July 2022. Different from digitizing the voice of a single person, producing the AI choir raises the problem of the trade-off between the convergence and diversity of the generated singing voices.

To produce a satisfactory choral effect, real singers must perform in a highly coordinated manner in terms of timbre, rhythm, and expressiveness. It is worth noting that the combination of the same voices does not produce a choral effect, and the choral effect actually results from the carefully crafted diversity of timbre, rhythm, and expressiveness of each singer in the collective performance. This yields a control problem for the joint generation of multiple singing voices: the coherence of the singers' voices and the diversity of timbres in the choir must be jointly optimized.

\begin{figure}
 \centering
 \includegraphics[width=7cm]{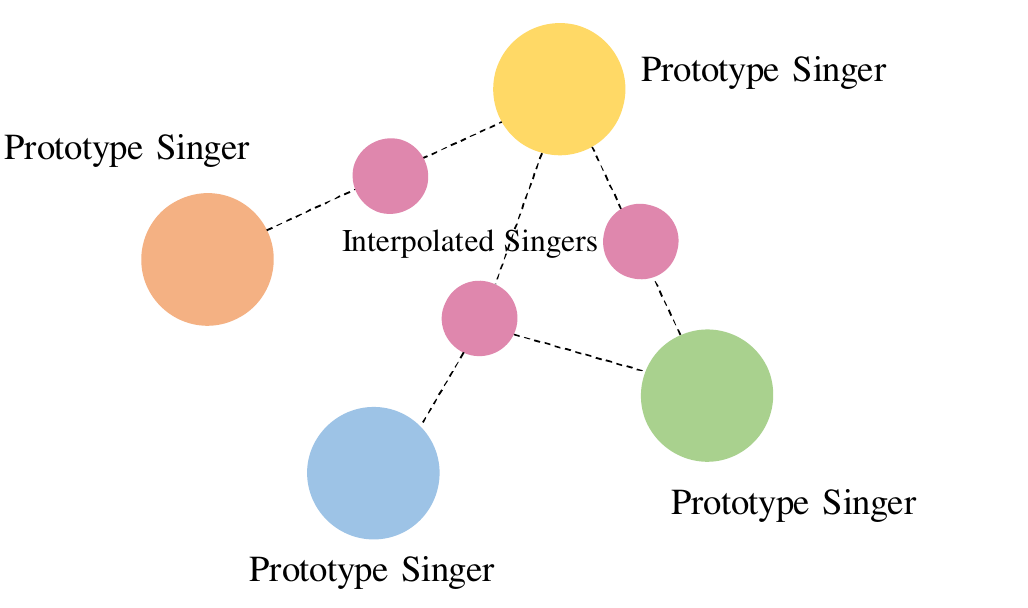}
 \caption{Generating interpolated virtual singers from prototype singers.}
 \label{fig5:proto}
\end{figure}

We develop a two-stage method to produce the AI choir with hundreds of virtual singers. First, several ``prototype'' singers with similar timbres are produced in the first stage. Then, in the second stage, as shown in Fig.~\ref{fig5:proto}, hundreds of new virtual singers are produced by interpolating between the timbres of these prototype singers. The harmonic choral effect is obtained by carefully controlling the rhythm and pitch of each virtual singer.

\begin{figure}
 \centering
 \includegraphics[width=11cm]{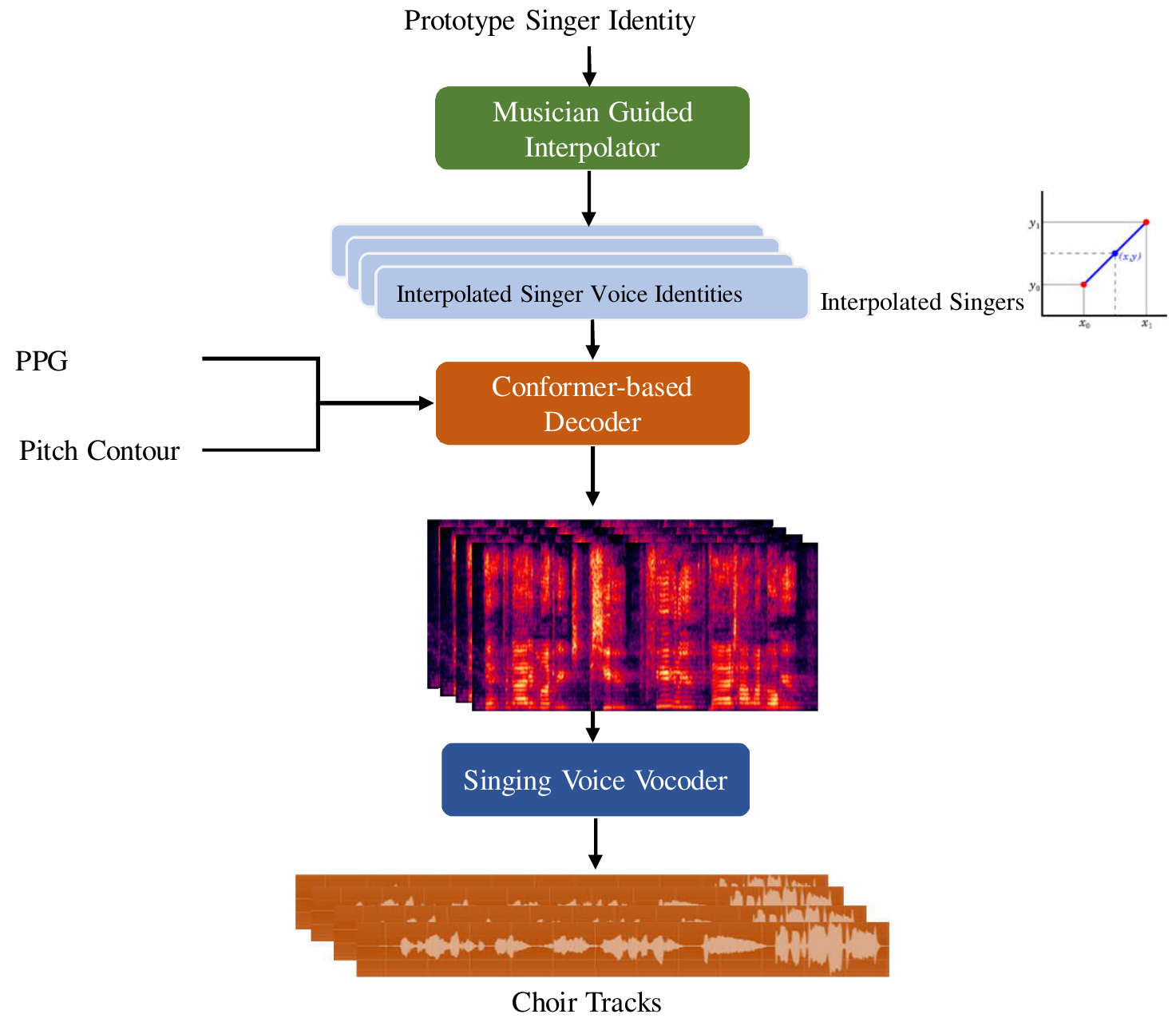}
 \caption{AI choir generation with timbre interpolation.}
 \label{fig6:interplation}
\end{figure}

\subsection{Prototype Singers}
The prototype singer is generated by digitizing the voices of an existing singer, given the clean vocal recordings. In practice, although many public-domain songs of a singer are available online, the vocal signals are mixed with accompaniments. To collect the large-scale training data, a pre-trained speech separation model, Demucs \cite{Defossez2019}, is used to extract the vocals from the original sounds. With the extracted vocal recordings from multiple prototype singers, we train a general model which relies on the speaker embedding to control the timbre of the produced singing voice. To produce the choir, eight prototype singers whose similar timbres are carefully selected by human evaluators are generated.

\subsection{Interpolated Singers}

The prototype singers can be seen as audio-form digital twins of existing singers, and new virtual singers can be created by performing timbre interpolation. As shown in Fig.~\ref{fig6:interplation}, speaker embeddings of different new virtual singers are generated by performing linear interpolation among the prototype singers, and these embeddings are utilized to produce the mel-spectrograms.

The same PPG and pitch contour embeddings are used for all different virtual singers, so the produced singing voices have the same lyrics and rhythm. The generated mel-spectrograms are finally converted to audio waveforms using the universal vocoder trained on datasets combining all singers. In total, 320 Virtual singers are produced to finally generate a choir.

Since prototype singers with similar timbres are used, although the timbres of all interpolated virtual singers are different, they will still be similar, which ensures collaborative performances in the choir. The human evaluator also plays an essential role in combining the virtual singers to produce the choir. The combination of a large number of virtual singers to produce the choir appears to be straightforward. However, human evaluators must examine whether the choir has a good combined timbre and adjust the proportions of the prototype singers.

\section{Experiments}
\label{exp}

\subsection{Datasets}
Three different datasets are used to evaluate the performances of singing voice generation, speaking-to-singing, and AI choir generation, respectively.

\textbf{OpenCpop.} The publicly-available high-quality Mandarin singing corpus, OpenCpop \cite{Wang2022}, is adopted to examine the capability of the proposed method to generate high-quality singing audio. The corpus consists of $100$ songs without accompaniments recorded by a professional female singer, and the audios are segmented into 3,756 utterances with a total duration of $5.2$~hours. Although the note and phoneme information is included in the original dataset, the proposed method uses only the audio waveforms to digitize the singer's voice.

\textbf{Speaking Audios.} To test the speaking-to-singing performances, we constructed a $3.7$~hours speaking audio dataset by ourselves based on the recordings of one male colleague in the daily Zoom meetings. We note that many high-quality publicly-available datasets (e.g., LibriTTS \cite{Zen2019}) for the reading speech synthesis are available. However, using the data collected from Zoom meetings helps to examine the feasibility of using daily normal-quality speeches to achieve voice digitization. To facilitate discussion, we denote this dataset as ``Speaking'' in the rest of the experiments.

\textbf{Audios from Multiple Singers.} To produce the choir, we further collected the songs of eight professional singers online (Youtube, Spotify, etc.), with four male singers and four female singers included. For each singer, nearly $4$ hours of data are collected, and for all songs, Demucs \cite{Defossez2019} is used to extract the vocal tracks. We denote this dataset as ``Multi-Sing'' in the following.

For all datasets, the audios are converted into the single-channel with a $24$~kHz sampling rate. No other information is required to conduct the unsupervised training. The obtained models are tested by synthesizing singing voices according to the ``straight'' excerpts of the VocalSet \cite{Wilkins2018}, with matched genders for the training dataset and testing utterances.

\subsection{Implementation}
\textbf{Acoustic model.} The structure of the acoustic model is shown in Fig.~\ref{fig3:conformer}. The dimensions of PPG, speaker identity and pitch contour embeddings are $320$, $256$, and $1$, respectively, and they all are transformed by linear layers to the dimension of $320$. $6$ Conformer blocks are included in the acoustic model, and for each Conformer block, we use $2$ attention heads. $80$-dimension mel-spectrogram is used as the training target, and the analysis frame size is $25$~ms with $10$~ms hop size. The maximum length of the Conformer is $1000$ frames, corresponding to $10$~s for the $10$~ms hop size. The model is trained with the batch size as $24$ with the learning rate as $1\times10^{-3}$ and the weight decay as $1\times10^{-6}$. The training is conducted for $2000$ epochs.

The resulting model is speaker-dependent for the OpenCpop and Speaking datasets since only one person is included in the training dataset. When training on the Multi-Sing dataset, the timbre of the output audio can be controlled by the speaker identity embedding.

\textbf{Vocoder.} To produce audio waveforms at $24$~kHz sampling rate from the $80$-dimension mel-spectrogram, $4$ transposed convolution-based upsampling blocks are included in the HiFi-GAN generator, with upsampling rates as $\{8,8,2,2\}$ and upsampling kernel sizes as $\{16,16,4,4\}$, respectively. The segment size for sequence-to-sequence modeling is 8192 samples. We used the batch size as $32$ for training with the learning rate as $2\times10^{-4}$, and the model was trained for $200,000$ steps.

Similar to the acoustic model, speaker-dependent models are obtained for the OpenCpop and Speaking datasets, and a universal vocoder is trained by using all the data in the Multi-Sing dataset.

\textbf{End-to-end Training.} After training the acoustic model and vocoder for each dataset, we finally perform end-to-end training to finetune both models further. We first freeze the acoustic model and update the vocoder for $5,000$ steps, then freeze the resulting vocoder to update the acoustic model for $100$ epochs. Such process is repeated 5 times.

\begin{figure} [t]
 \centering
 \includegraphics[width=10cm]{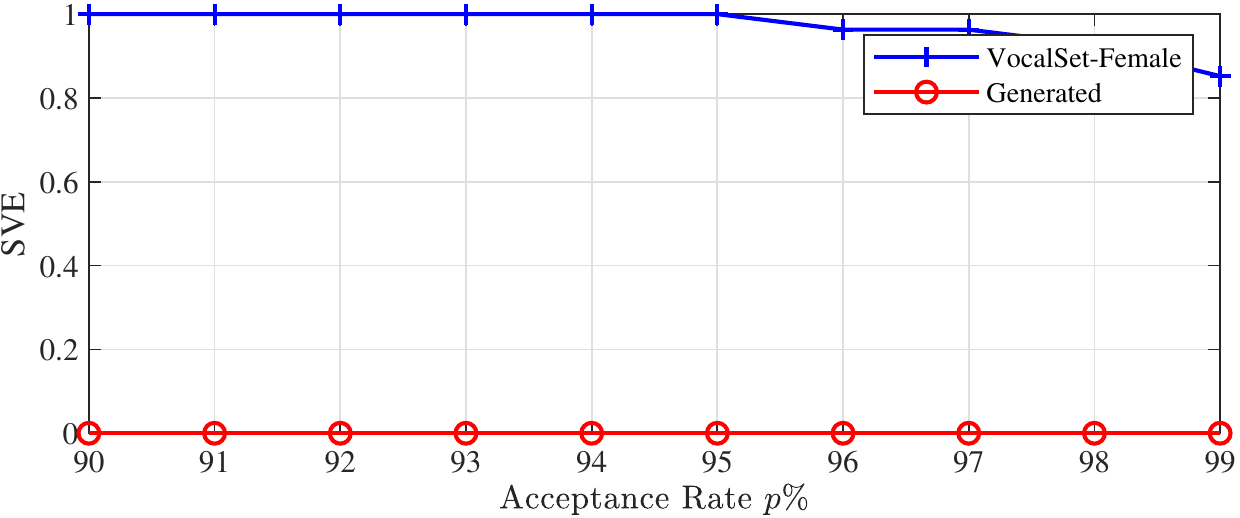}
 \caption{The SVE as a function of acceptance rate for the OpenCpop dataset. The acceptance rate is used to determine the threshold for speaker verification.}
 \label{fig7:p}
\end{figure}

\begin{figure} [t]
 \centering
 \includegraphics[width=13cm]{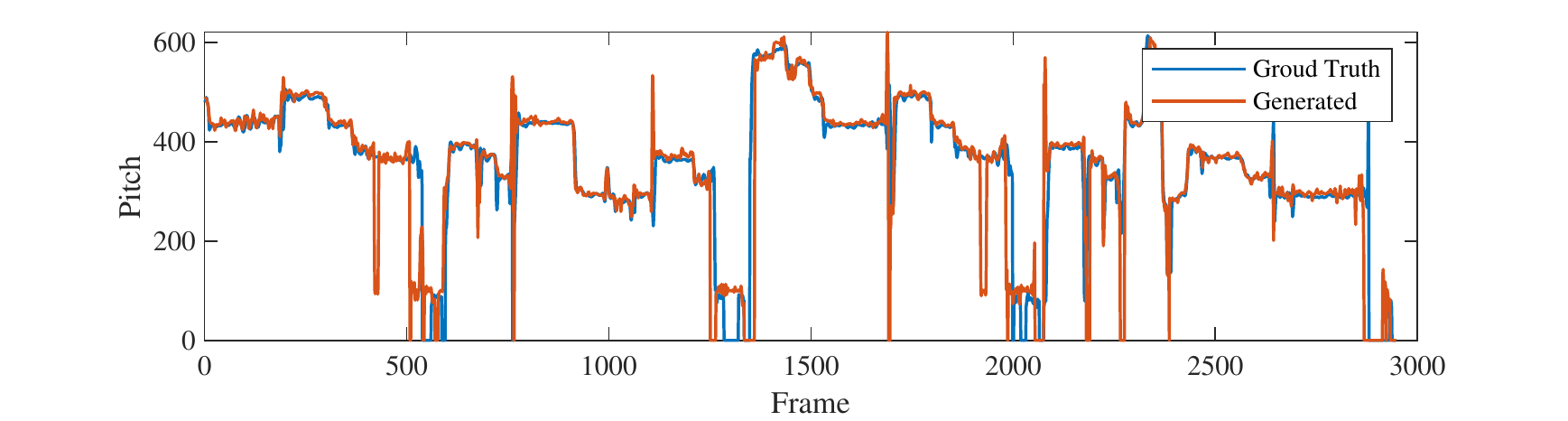}
 \caption{Comparison between the pitch contours in the generated audio and the ground truth for singing voice generated by the OpenCpop singer.}
 \label{fig8:pitch}
\end{figure}


\subsection{Results}

\textbf{Singing Voice Generation.} In this experiment, we digitize the OpenCpop singer based on her singing recordings.

We first check the speaker identity similarity of the produced audios. With the pre-trained ECAPA-TDNN model \cite{Desplanques2020}, the speaker embeddings of all utterances in the OpenCpop dataset are computed and then averaged to get the overall speaker embedding of OpenCpop singer, which is denoted by $\mathbf{S}_\textrm{cpop}$ here. In the following, the $\mathbf{S}_\textrm{cpop}$ will be used as an anchor to examine whether the produced audio is similar to the OpenCpop singer according to a threshold. For all utterances in the OpenCpop,  the cosine similarities between their speaker embeddings and $\mathbf{S}_\textrm{cpop}$ are computed, then the threshold is determined by accepting`` $p\%$'' utterances after sorting the similarity scores. It can be seen that the higher $p$, the lower the threshold.

\begin{figure}[t]
 \centering
 \includegraphics[width=10cm]{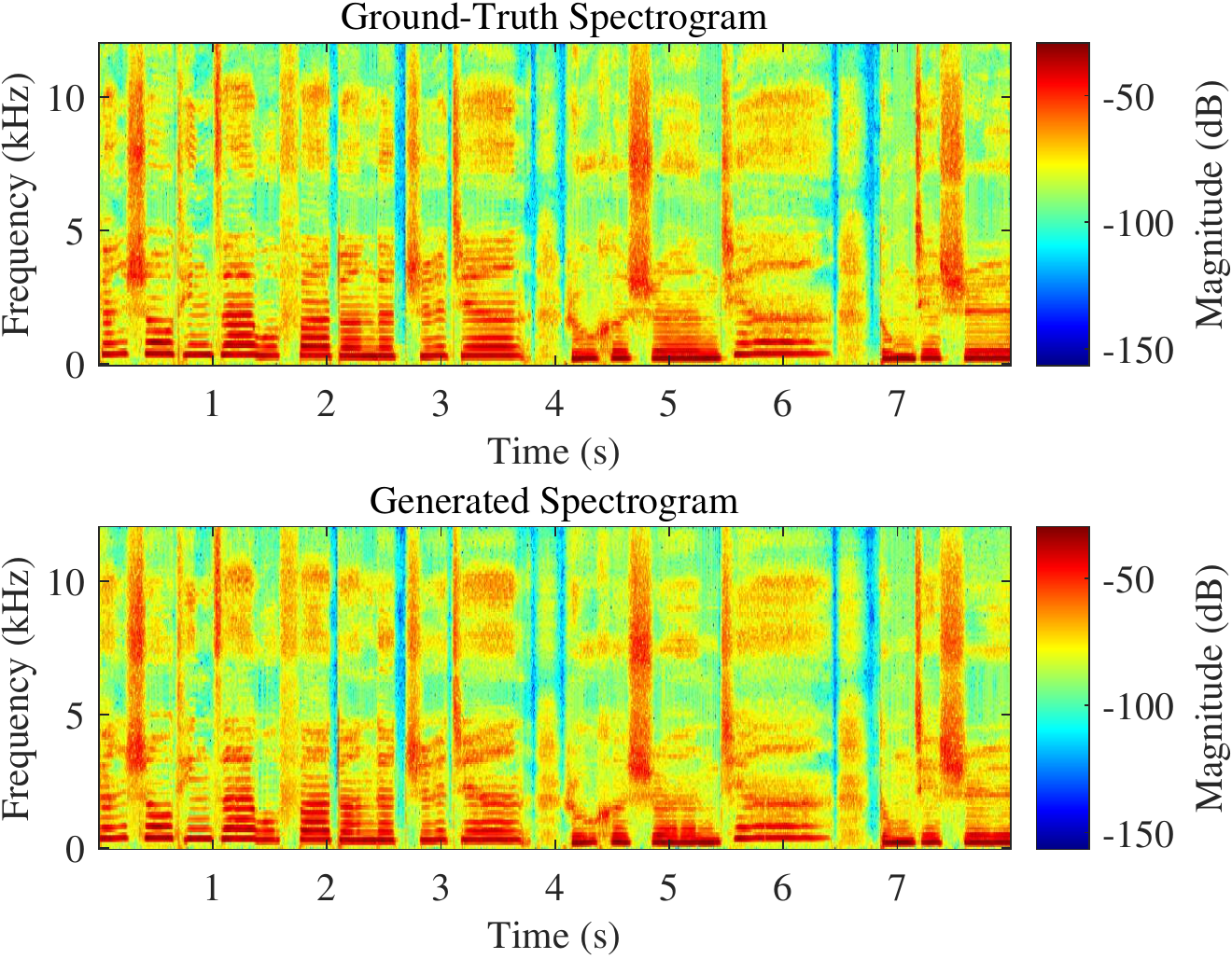}
 \caption{Linear spectrograms of the ground truth and generated signal for the OpenCpop singer.}
 \label{fig9:spec}
\end{figure}

\begin{figure}[t]
 \centering
 \includegraphics[width=10cm]{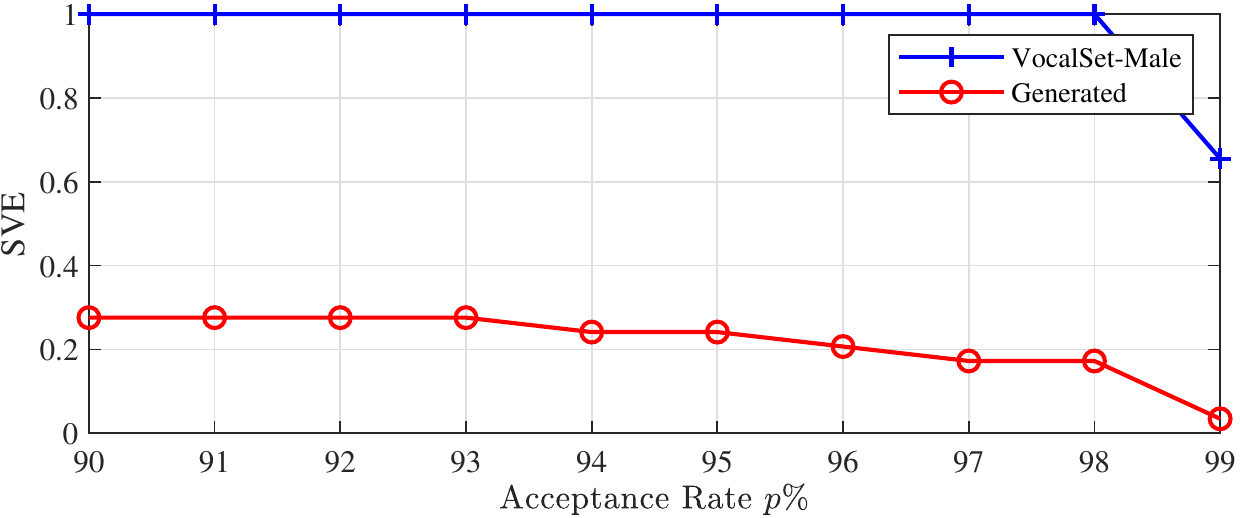}
 \caption{The SVE as a function of acceptance rate for the ``Speaking'' dataset. The acceptance rate is used to determine the threshold for speaker verification.}
 \label{fig10:guo_p}
\end{figure}

As described above, 25 utterances are generated by using the 25 ``straight'' excerpts of the VocalSet from the female singers. The cosine similarities between the generated audios and $\mathbf{S}_\textrm{cpop}$ are computed, and the Speaker Verification Error (SVE) is obtained by comparing the similarities with the threshold. For comparison, the same metric for the original excerpts of the VocalSet is also calculated. Fig.~\ref{fig7:p} shows the SVE of the generated audios and the VocalSet audios as a function of the acceptance rate $p\%$. For all acceptance rates, the SVE of the generated audios is zero compared to the VocalSet audios having SVE near to one, which indicates that the proposed method can effectively learn the timbre of the OpenCpop singer.

Then we evaluate the pitch accuracy of the generated audios. Fig.~\ref{fig8:pitch} illustrates an example of the pitch contour of the generated audio and the pitch contour of the input audio from the VocalSet. By taking the pitch contour as the controlling factor of the acoustic model, the generated audio is expected to follow the specified melody. It is shown that in almost all cases, the generated audio produces the desired pitches, except for some errors on onset and offset.

We further show one example of the spectrograms of the generated audio and the ground truth in Fig.~\ref{fig9:spec}. The generated audio is obtained by first extracting the embeddings from the ground truth audio and then using the unaltered embeddings to reconstruct the waveform signal. We can observe that the generated audio can effectively capture the temporal-frequency characteristics of the original singing audio, indicating the capability to produce high-quality audio outputs.

\begin{figure}[t]
 \centering
 \includegraphics[width=13cm]{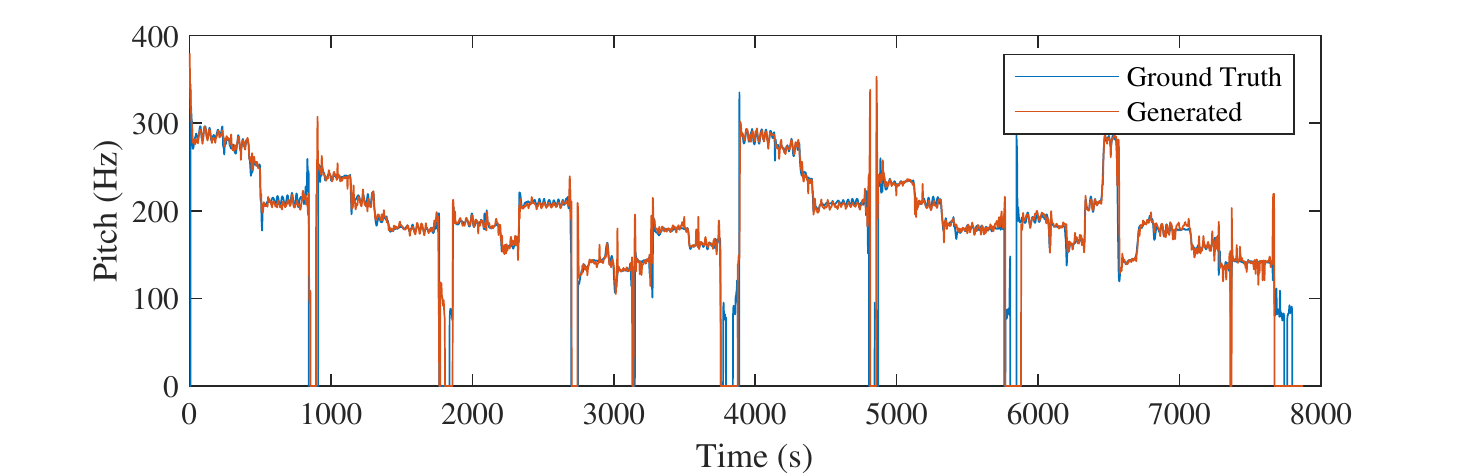}
 \caption{Comparison between the singing pitch contours in the generated audio and the ground truth for speaking to singing by a male speaker.}
 \label{fig11:pitch_singing_guo}
\end{figure}

\begin{figure}[t]
 \centering
 \includegraphics[width=10cm]{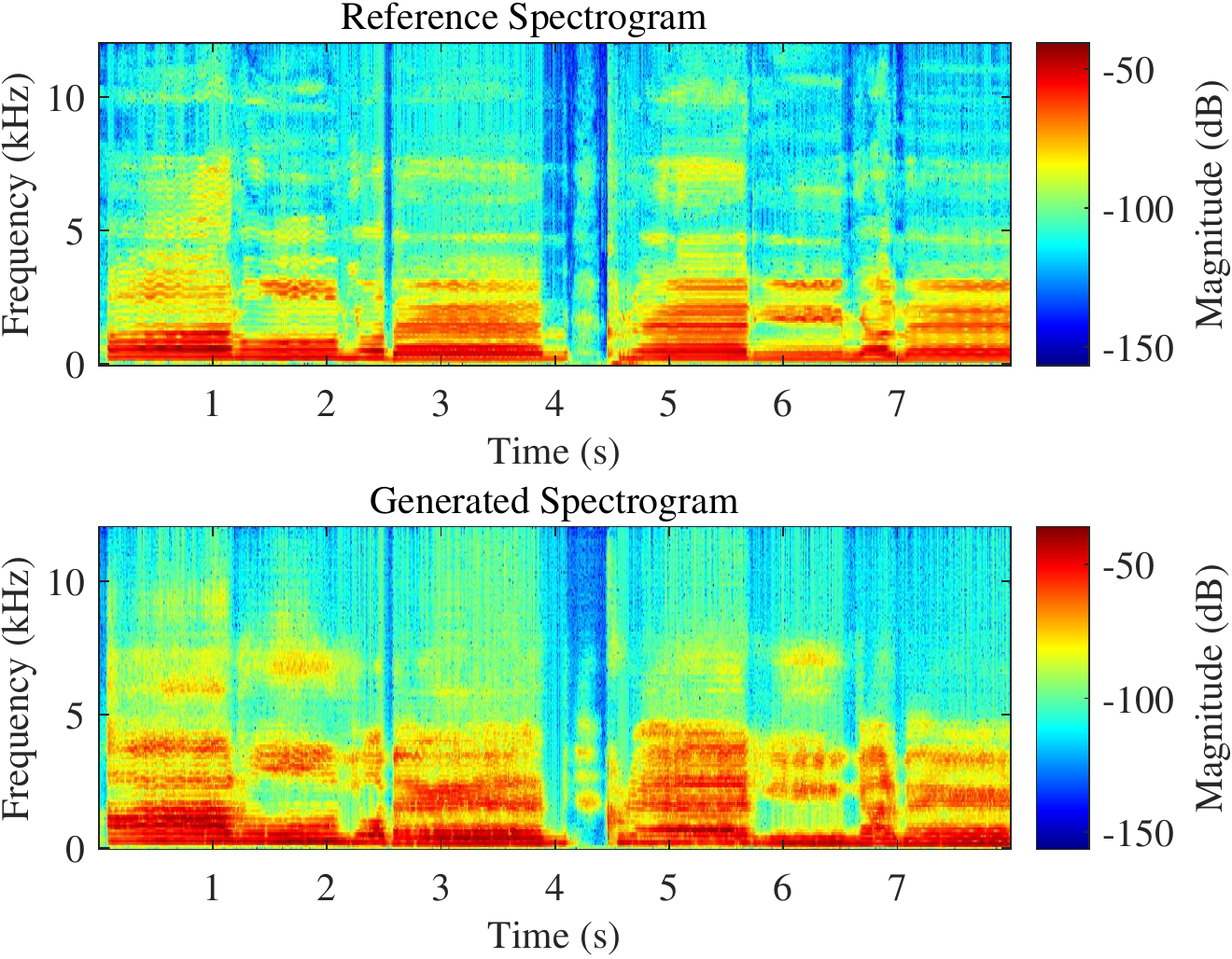}
 \caption{Linear spectrograms of the reference and generated signal for the ``Speaking'' dataset. A singing voice is produced. We note that the reference and generated audios correspond to different speakers.}
 \label{fig12:spec_singing_guo}
\end{figure}

\textbf{Speaking to Singing.} Using the ``Speaking'' dataset, we further evaluate whether the proposed method could generate singing audios based on the dataset only containing the speaking voices. Similar to the experiments for the OpenCpop dataset, we use the $29$ ``straight'' excerpts of the VocalSet from the male singers as the input signals to the models. The results are shown in Fig.~\ref{fig10:guo_p}. We can see that when rejecting $10\%$ training data according to the speaker embedding similarity, nearly $27\%$ generated audios are classified as different from the target speaker. By increasing the acceptance rate to  $99\%$, almost all generated samples are regarded as the target, indicating that the largest divergence of the generated audios to the average speaker embedding is comparable to the training set. We also notice that the SVE values for all cases are apparently higher than the OpenCpop cases, showing that generating the singing audios from speaking-only data is more challenging.

\begin{figure}[t]
 \centering
 \includegraphics[width=10cm]{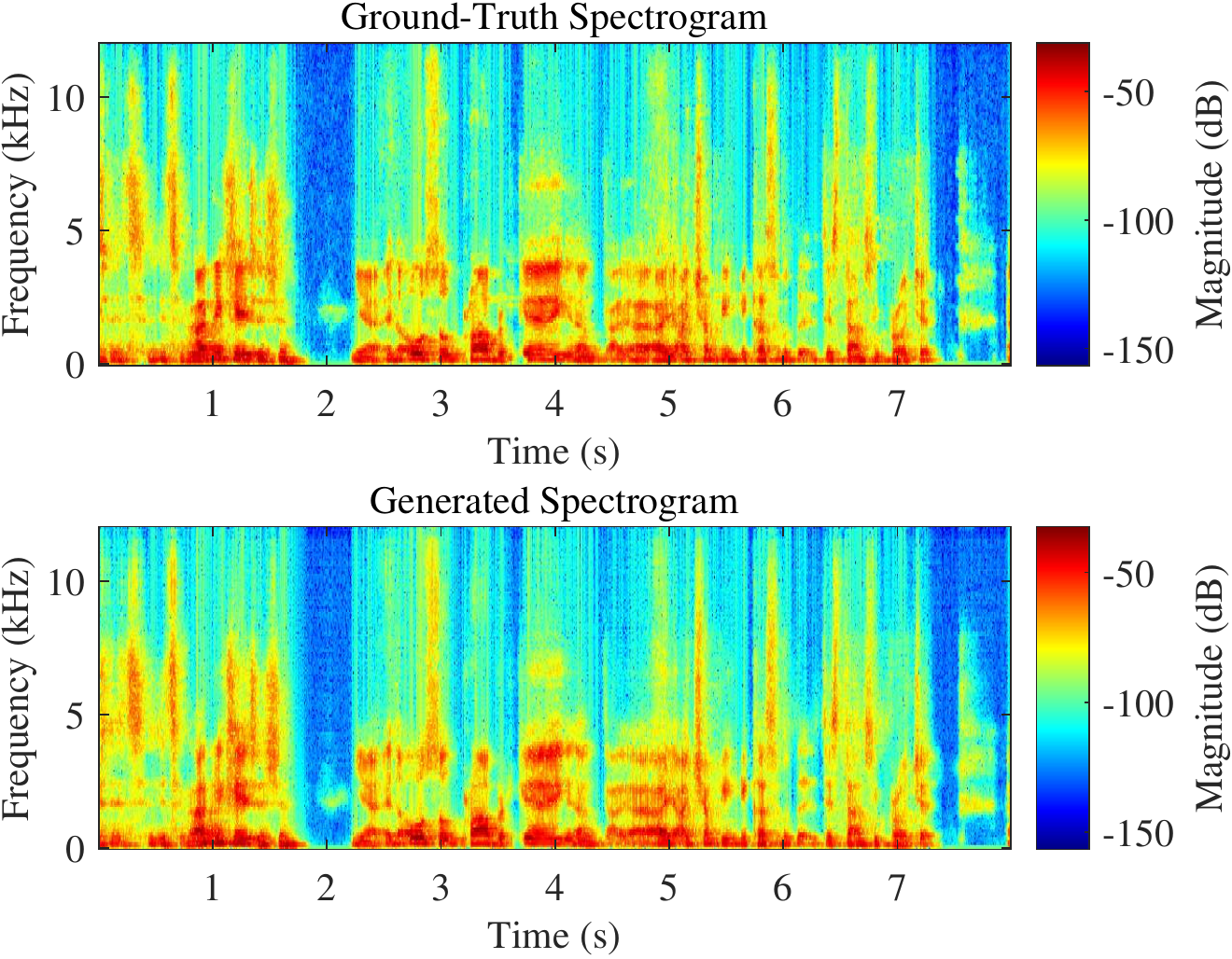}
 \caption{Linear spectrograms of the ground-truth and generated signal for the ``Speaking'' dataset. Speaking voice is reconstructed. We note that the ground truth and generated audios correspond to the same speaker.}
 \label{fig13:spec_guo}
\end{figure}

\begin{figure}[t]
 \centering
 \includegraphics[width=10cm]{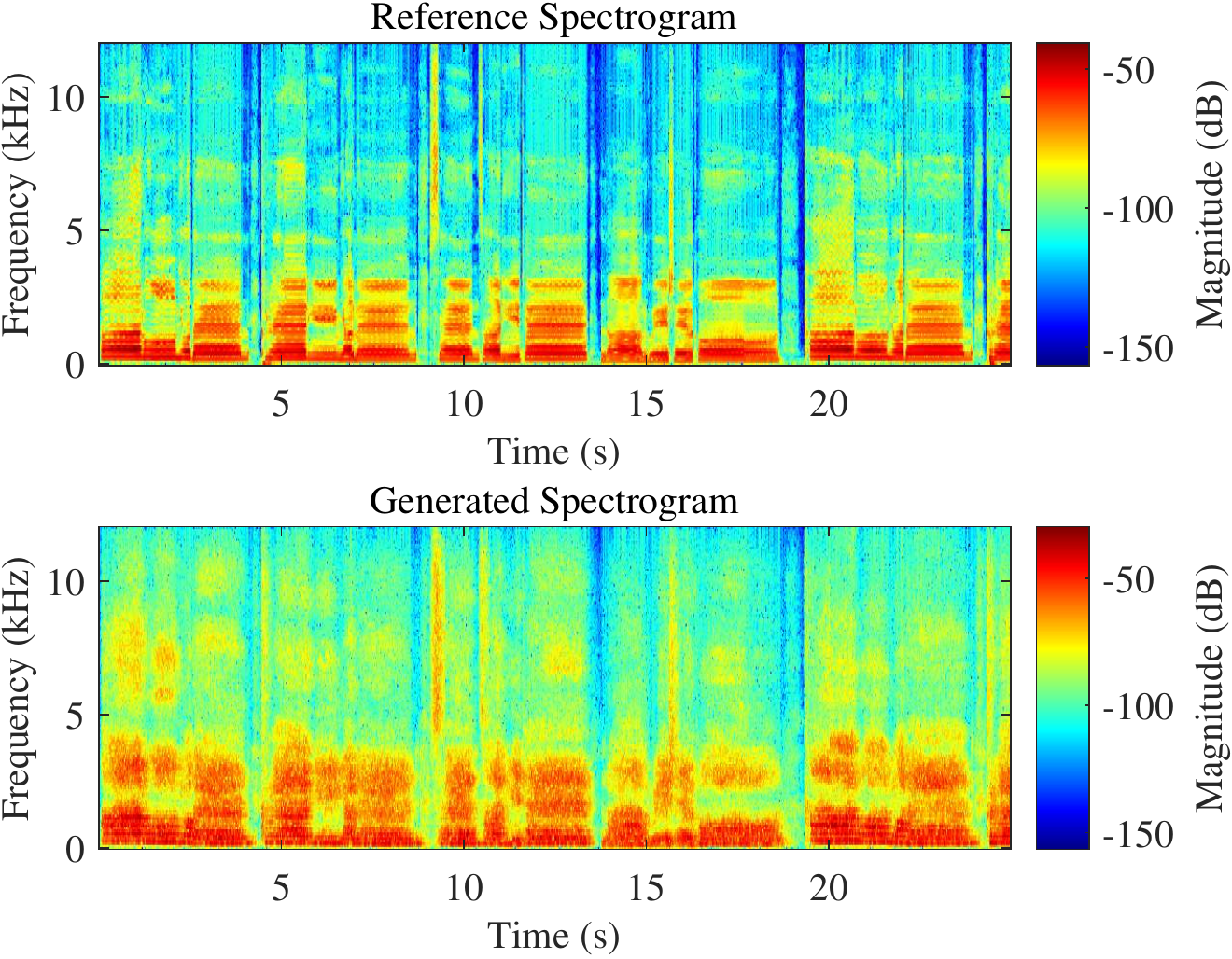}
 \caption{Linear spectrograms of the reference and a generated choir consisting of 120 male singers.}
 \label{fig14:choir_male}
\end{figure}

The pitch contours of the generated singing audio and ground truth are compared in Fig.~\ref{fig11:pitch_singing_guo}. We can see that even though the model is trained on the speaking-only dataset, the resulting model can still produce singing pitch contours with high precision. In Fig.~\ref{fig12:spec_singing_guo}, we compare the spectrograms of the reference audio from the VocalSet and the generated audio. It can be noticed that although the produced signal can effectively follow the melody and rhythm of the reference signal, their frequency distributions are different. This, in turn, shows that the produced audio has exhibited a different formant from the reference one, indicating a successful modification of the speaker identity. The proposed method can achieve accurate reconstruction performance simply based on the extracted embeddings when reconstructing the speaking audio from the training speaker, as shown in Fig.~\ref{fig13:spec_guo}.

\begin{figure}[h]
 \centering
 \includegraphics[width=10cm]{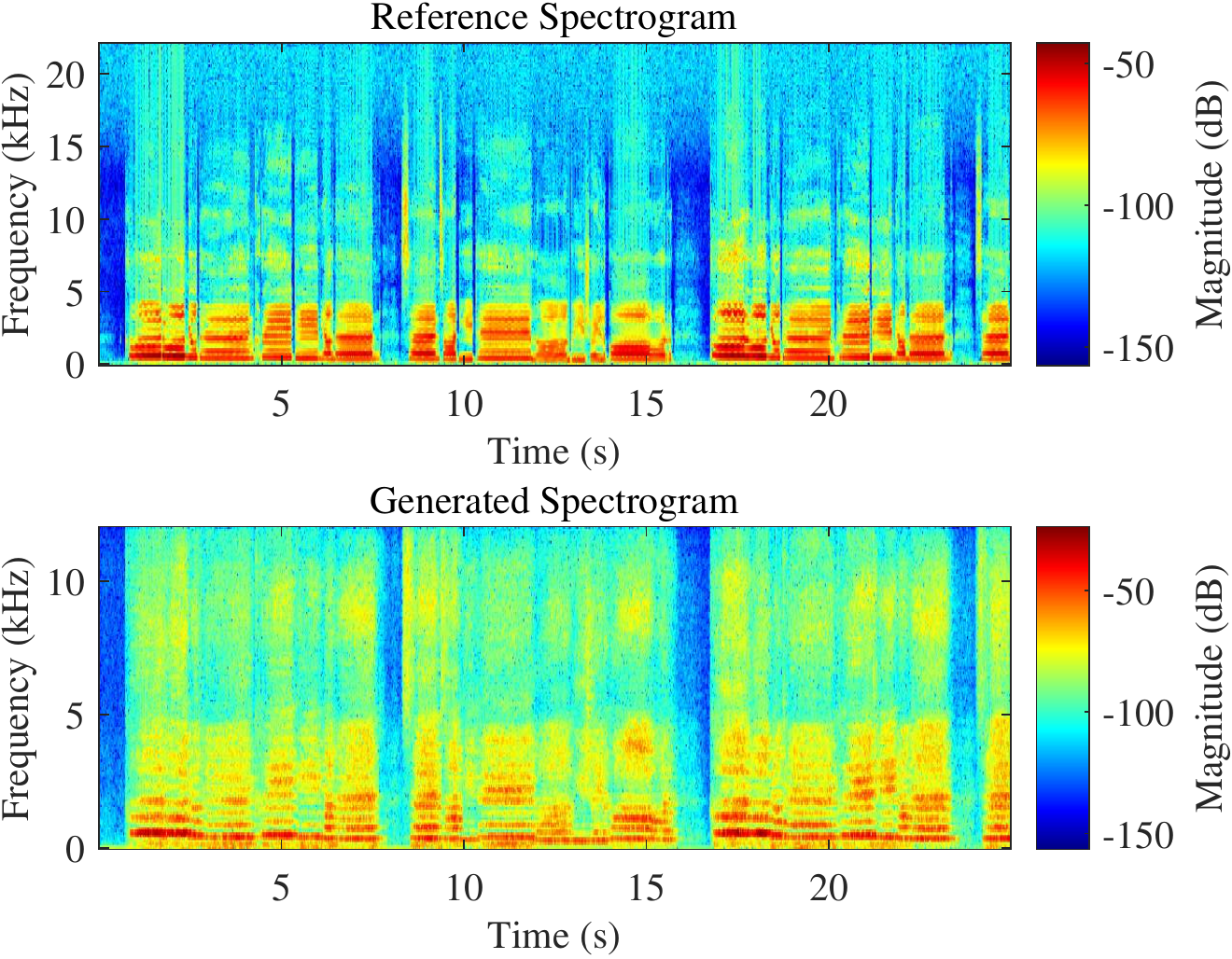}
 \caption{Linear spectrograms of the reference and a generated choir consisting of 120 female singers.}
 \label{fig15:choir_female}
\end{figure}

\textbf{Choir Generation.} We finally evaluate the performance of the proposed method of generating the choir, which consists of tens to hundreds of virtual singers. By using identity interpolation, new virtual singers can be produced from the prototype singers, who are trained using the data collected online. We note that the singing voice performances for each singer are similar to the OpenCpop singer, and the same conclusions can be drawn; therefore, evaluations on each generated audio are not presented here. In Fig.~\ref{fig14:choir_male}, we show an AI choir consisting of 160 male virtual singers. We notice that since each singer has a unique formant, the pitch harmonics become indistinguishable in the combined choir signal, which leads to the choir effect. Similar results can be observed from the female choir shown in Fig.~\ref{fig15:choir_female}, where the frequency distribution is much broader than the male choir due to the higher diversity of female timbres.

\section{Conclusion}
This paper presents a novel framework for unsupervised voice modeling that enables the creation of digital singing humans. Our method relies on a variational auto-encoder (VAE) that encodes audio recordings as various hidden embeddings representing different factors of the singing voice, which can then be manipulated to control various singing characteristics such as pitch, melody, and lyrics. By training on large-scale data, the proposed method can also learn to model other unique skills, including the accent and emotional expression of the singer. Experimental results on different datasets demonstrate the effectiveness of the proposed method.

\medskip

\newpage
\small
\bibliographystyle{IEEEtran}
\bibliography{IEEEabrv,sampref}

\end{document}